\documentclass{ws-ijmpcs}

\usepackage{epsfig}
\usepackage{graphicx}
\usepackage{slashed}
\usepackage{hyperref}

\newcommand{\be}{\begin{equation}}
\newcommand{\ee}{\end{equation}}
\newcommand{\bea}{\begin{eqnarray}}
\newcommand{\eea}{\end{eqnarray}}

\newcommand{\m}{{\scriptscriptstyle -}}

\def\lambdabar{\protect\@lambdabar}
\def\@lambdabar{%
\relax \bgroup
\def\@tempa{\hbox{\raise.73\ht0
\hbox to0pt{\kern.2\wd0\vrule width.7\wd0
height.1pt depth.1pt\hss}\box0}}%
\mathchoice{\setbox0\hbox{$\displaystyle\lambda$}\@tempa}%
{\setbox0\hbox{$\textstyle\lambda$}\@tempa}%
{\setbox0\hbox{$\scriptstyle\lambda$}\@tempa}%
{\setbox0\hbox{$\scriptscriptstyle\lambda$}\@tempa}%
\egroup }

\begin{document}

\markboth{T.\ Heinzl}{Strong-Field QED and High Power Lasers}

%
\catchline{}{}{}{}{}
%

\title{STRONG-FIELD QED AND HIGH POWER LASERS
}

\author{THOMAS HEINZL
}

\address{School of Computing and Mathematics, Plymouth University, Drake Circus\\
Plymouth, PL4 8AA, UK
\\
theinzl@plymouth.ac.uk}

%

\maketitle

\begin{history}
\end{history}

\begin{abstract}
This contribution presents an overview of fundamental QED processes in the presence of an external field produced by an ultra-intense laser. The discussion focusses on the basic intensity effects on vacuum polarisation and the prospects for their observation. Some historical remarks are added where appropriate.

\end{abstract}

\ccode{PACS numbers: 12.20.-m, 42.50.Xa}

\section{Introduction}	

It is a time honoured principle of physics research that new discoveries are typically made upon exploring new parameter regimes. Arguably, the most important parameter is energy which governs the experimental `resolution' from the everyday macroworld down to the microworld of high energy physics. Other relevant parameters that may be tuned at will include temperature, density, pressure etc. This conference has seen many talks where geometry plays a crucial role in the guise of boundary conditions governing the Casimir effect. Finally, one may consider to vary the magnitude of an external field which, in its magnetic incarnation, for instance, can have drastic effects on the physics of spin systems and superconductors. In this presentation I will concentrate on the peculiar external fields provided by ultra-intense, high-power lasers\footnote{For a topical and detailed review including a wealth of references see Ref.~\refcite{DiPiazza:2011tq}.}

\begin{figure}
\begin{center}
  \includegraphics[angle=90,scale=0.4]{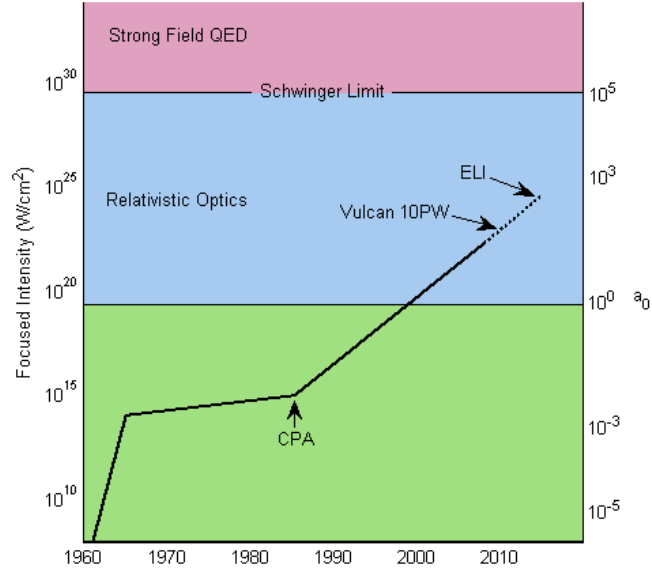}
\end{center}
\caption{Historical development of laser intensity (adapted from Ref.~\protect\refcite{tajima:2002}). CPA denotes the breakthrough provided by `chirped pulse amplification'\protect\cite{strickland:1985}. \label{fig:laserIntensity}}
\end{figure}

Traditionally, the strength of the laser field is measured by the dimensionless laser amplitude,
\be \label{A0}
  a_0 \equiv \frac{e E \lambdabar_L}{mc^2} \; ,
\ee
which measures the energy gain of an electron (charge $e$, mass $m$) traversing a laser wavelength $\lambdabar_L$ in the r.m.s.\ field $E$ in units of its rest energy, $mc^2$. Clearly, when $a_0$ exceeds unity, the electron becomes relativistic. If we measure intensity $I$ by the modulus of the Poynting vector ($I = c E^2$ for a plane wave) and denote the current record intensity\cite{Yanovsky:2008} by $I_{22} \equiv 10^{22}$ W/cm$^2$ we can relate $a_0 \simeq 60 \sqrt{I/I_{22}} \, \lambda/\mu\mbox{m}$. Hence, one may expect $a_0$ values of order 200 at Vulcan 10 Petawatt\cite{Vulcan10PW:2009} and $10^3$ at ELI\cite{ELI:2009} in the not-too-distant future. The historical development of laser intensity is shown in Fig.~\ref{fig:laserIntensity}, and Table~\ref{table:MAGNITUDES} provides an overview of the (current) magnitudes involved.

\begin{table}[h]
\tbl{Some typical magnitudes characterising current high-power lasers.}
{\begin{tabular}{ll} \toprule
Quantity & Magnitude\\ \colrule
Power & $ P \gtrsim 10^{15} \, \mbox{W}  = 1 \, \mbox{PW}$ \\
Intensity & $I \, \gtrsim 10^{22} \, \mbox{W/cm}^2$ \\
Electric field & $E \gtrsim 10^{14} \, \mbox{V/m}$ \\
Magnetic field & $B \gtrsim 10^{10} \, \mbox{G} = 10^6 \, \mbox{T}$\\
\botrule
\end{tabular}\label{table:MAGNITUDES}}
\end{table}

Although the electric and magnetic fields are among the largest that can be generated in a lab they come with a price. Typically, they are extremely short-lived (pulse duration a few femtoseconds) and alternating in sign. Hence, though macroscopic, the fields differ vastly from homogeneous configurations. This will be further elucidated below.

\section{Strong Laser Fields: Theory}

\subsection{Modelling the laser beam}

The primary theory input is the model of the laser beam and/or pulse. The simplest possibility is to choose an infinite plane wave, that is, a field strength tensor $F^{\mu\nu}$ depending solely on the invariant phase, $\phi = K \cdot x$ where $K$ is the wave four-vector. The dependence on $\phi$ is arbitrary, but a natural example is a monochromatic beam with a sinusoidal dependence on $\phi$ which may be augmented by a pulse envelope. As long as the latter depends solely on $\phi$ we are still dealing with a plane wave. Clearly, the pulse duration (related to the width of the envelope) will then be an additional parameter. This field configuration is still somewhat unphysical as it has infinite transverse extent. To remedy this one may model the wave as a Gaussian beam which has a Gaussian profile in the transverse direction. Accordingly, one now has longitudinal \emph{and} transverse scales given by the Rayleigh length $z_R$ and the waist $w$, respectively. The dimensionless ratio\cite{Narozhny:2000},
\be \label{KAPPA}
  \kappa \equiv w/z_R \; ,
\ee
then measures the deviation from the plane wave case corresponding to $\kappa = 0$. In the opposite direction one may increase $\kappa$ until one reaches the diffraction limit (extreme focussing) corresponding to $\kappa = 1/2\pi$. This suggests that perturbation theory in $\kappa \ll 1$ might generally be a good idea. In the following, though, we will be mostly concerned with the plane wave case ($\kappa = 0$).

It is worthwhile to recall some peculiarities of plane waves from a covariant perspective. Their wave vector is of course null, $K^2 = 0$, and we have already seen that the field strength is a univariate function, $F^{\mu\nu} = F^{\mu\nu} (K \cdot x)$. The vector $K$ singles out a frame and we assume that the plane wave laser beam propagates along $z$ with frequency $\Omega$, i.e.\ $K = \Omega (1, \hat{\mathbf{z}})$. In this case, $K \cdot x = \Omega x^\m$, which tells us that the field depends solely on the light-front variable\cite{Heinzl:2000ht} $x^\m = ct - z$. Maxwell's equations in vacuum then demand that $F^{\mu\nu}$ is transverse, $K_\mu F^{\mu\nu} = 0$. Most importantly, a plane wave represents a \emph{null field} for which the basic invariants vanish\footnote{For Gaussian beams these turn out to be of order $\kappa^2$.},
\be \label{SP}
  \mathcal{S} \equiv - \frac{1}{4} F_{\mu\nu}F^{\mu\nu} = 0 \; , \quad \mathcal{P} \equiv - \frac{1}{4} F_{\mu\nu} \tilde{F}^{\mu\nu} = 0 \; .
\ee
In addition, the matrix cube of $F^{\mu\nu}$ vanishes as well, $F^3 = 0$, i.e.\ $F^{\mu\nu}$ is nilpotent with index 3. Due to these null field properties it is not straightforward to measure the strength of a plane wave field in an invariant way. What is required is a probe which provides an independent four-momentum, say a photon with wave vector $k = \omega (1, \mathbf{n})$. In this case, one can form two basic invariants, namely,
\bea
  I_1 &\equiv& k \cdot K \stackrel{\mathrm{lab}}{=} 2 \Omega \omega \; , \label{I1} \\
  I_2 &\equiv& k_\mu T^{\mu\nu} k_\nu \stackrel{\mathrm{lab}}{=} 2 \omega^2 E^2 \; . \label{I2}
\eea
Here we have assumed a head-on collision ($\mathbf{n} = - \hat{\mathbf{z}}$) in the lab frame for simplicity. The second invariant utilises the Maxwell energy momentum tensor, $T^{\mu\nu} = F^{\mu\lambda}F_\lambda^{\;\; \nu} = (E^2/\Omega^2)K^\mu K^\nu $, and may be regarded as the energy density `seen' by the probe. A null energy theorem\cite{Shore:2007um} says that $I_2$ has to be nonnegative which is nicely borne out.

For what follows it is useful to make the invariants dimensionless. To this end we recall Sauter's critical electric field\cite{Sauter:1931zz}, the `Schwinger limit' of Fig.~\ref{fig:laserIntensity},
\be \label{SAUTER}
  E_S \equiv \frac{m^2}{e} \simeq 1.3 \times 10^{18} \; \mbox{V/m} \; ,
\ee
setting $\hbar = c = 1$ henceforth. In such a field an electron gains an energy $m$ across a Compton wavelength, $1/m$, and hence the probability of vacuum pair production becomes sizeable\footnote{It should be  emphasized at this point that the field (\ref{SAUTER}) has a magnitude typical of any QED process. The difficulty resides in creating such a field across a \emph{macroscopic} distance (say a laser focus) and not just over a Compton wavelength.}. In QED it thus makes sense to measure electric fields in units of the Sauter field (\ref{SAUTER}) resulting in the dimensionless field strength parameter,
\be
  \epsilon \equiv E/E_S \; .
\ee
Similarly we measure laser and probe frequencies in terms of the electron rest mass, $m$,
\be \label{NU}
  \nu_L \equiv \Omega/m \; , \quad \nu \equiv \omega/m \; .
\ee
In terms of the invariants (\ref{I1}) and (\ref{I2}) we thus find the relations
\bea
  I_1/m^2 &\stackrel{\mathrm{lab}}{=}& 2 \nu \nu_L \; , \\
  e I_2^{1/2} /m^3 &\stackrel{\mathrm{lab}}{=}& 2 \epsilon \nu \; , \\
  a_0 = eI_2^{1/2} / mI_1 &\stackrel{\mathrm{lab}}{=}& \epsilon/\nu_L \label{A0.INV} \; .
\eea
Interestingly, the last identity (\ref{A0.INV}) provides an invariant definition of the dimensionless laser amplitude\cite{Heinzl:2008rh}.

\subsection{Strong field QED}

The model chosen for the laser beam enters the QED action via its classical \emph{external} gauge potential $A_\mu$. This becomes particularly obvious in the partition function
\be \label{Z}
  Z[A] = \int Da \, D\psi \, D\bar{\psi} \, \exp i \left\{ S_{\mathrm{QED}} [a, \psi , \bar{\psi}] - e (A, j) \right\} \; ,
\ee
where $S_{\mathrm{QED}}$ is the standard QED action depending on the the fermion fields $\psi$ and $\bar{\psi}$ and the fluctuating quantum field $a_\mu$ integrated over in the path integral. The external field $A_\mu$, however, does not have a dynamics of its own, its only role being the coupling to the fermionic current $j^\mu = \bar{\psi} \gamma^\mu \psi$. In other words by ignoring its kinetic term we are making the \emph{approximation} that there is \emph{no classical backreaction} of the current on the external field. The \emph{quantum} action, however, will have terms coupling $a$ and $A$, so that their dynamics becomes entangled (see below).

The S-matrix and Feynman diagrams for the theory (\ref{Z}) are then generated by absorbing the external field interaction into the unperturbed Hamiltonian and using the eigenstates of the latter as the asymptotic states. This Furry picture\cite{Furry:1951zz} requires the solution of the Dirac equation in the presence of the background field. Whether or not this is possible depends crucially on the choice of the laser beam model. For plane waves an analytic solution was found long ago by Volkov\cite{Wolkow:1935zz}. Integrability here is guaranteed by the fact that a plane wave is univariate so that a sufficient number of momentum components is conserved. As soon as the background becomes multivariate, this property is lost. Hence, in what follows we will only discuss the plane wave case resulting in the Volkov solution.

\begin{figure}[h]
\begin{center}
  \includegraphics[width = 0.9\textwidth]{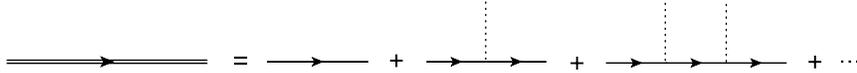}
\end{center}
\caption{Dressed fermion line expanded in terms of bare ones. The external plane wave field is given by the dotted lines representing continuous absorption/emission of laser photons. \label{fig:dressedProp}}
\end{figure}

The Feynman rules then are as follows. The quantum field $a_\mu$ corresponds to non-laser (probe) photons and will be described by standard wavy photon lines. The background field (dotted lines) will normally not be displayed explicitly but, through the Volkov solution, serves to dress the fermions which are represented by double lines, see Fig.~\ref{fig:dressedProp}.

\vspace{-.5cm}

\begin{figure}[h]
\begin{center}
  \includegraphics[width = 0.35\textwidth]{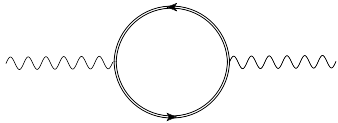}
\end{center}
\caption{Strong field vacuum polarisation diagram with two external non-laser photons. \label{fig:VacPol}}
\end{figure}

As usual, these fermion lines can be either external (asymptotic Volkov states) or internal (Volov propagator). With this `lego kit' of particle lines we can then form Feynman diagrams representing the `elementary' processes of laser QED. Tree diagrams (nonlinear Compton scattering, photon induced and trident pair production) have been covered in the talk by Anton Ilderton\cite{Ilderton:2011}. We will thus focus on fermion loop diagrams describing variants of vacuum polarisation processes such as displayed in Fig.~\ref{fig:VacPol}.

\begin{figure}[h]
\begin{center}
  \includegraphics[width = \textwidth]{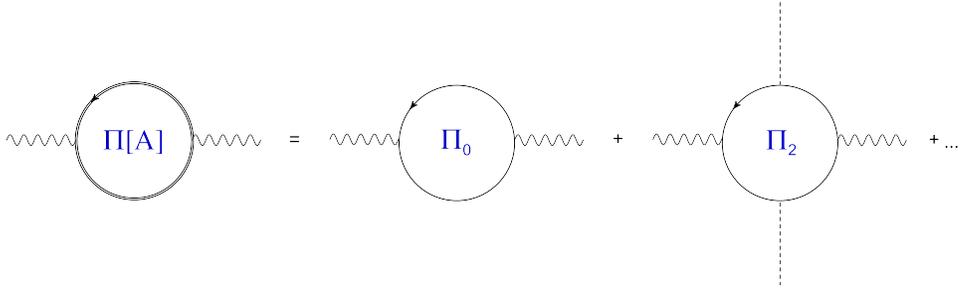}
\end{center}
\caption{The polarisation tensor expanded to lowest nontrivial order.\label{fig:VacPolExpanded}}
\end{figure}

The dressed fermion loop represents the vacuum polarisation tensor, $\Pi^{\mu\nu}[A]$ in the background of the plane wave field, $A_\mu$. Expanding this in powers of the external field (cf.\ Fig.~\ref{fig:dressedProp}) results in the standard QED term plus a leading order correction with four external photon lines (two probe and two background photons) as displayed in Fig.~\ref{fig:VacPolExpanded}. At low probe energies ($\omega \ll mc^2$) this latter diagram has first been worked out by Euler and Kockel\cite{Euler:1935zz}, two students of Heisenberg's.

In modern parlance their findings may be used to describe the second order term of Fig.~\ref{fig:VacPolExpanded} by an effective Lagrangian of the form
\be
  \Delta \mathcal{L}_{\mathrm{eff}} = \frac{1}{2} a_\mu \Pi_2^{\mu\nu}[A] a_\nu = \frac{2}{45}\frac{\alpha^2}{m^4}  \left(c_1 \mathcal{S}^2 + c_2 \mathcal{P}^2 \right) \; , \quad c_1 = 4 \; , \; c_2 = 7 \; .
\ee
On the right-hand side, $\mathcal{S}$ and $\mathcal{P}$ refer to to the invariants (\ref{SP}) with the replacement $F_{\mu\nu} \to F_{\mu\nu} + f_{\mu\nu}$ and discarding terms higher than second order in $F$ and $f$.  The main technical achievement by Euler and Kockel was the determination of the coefficients $c_1$ and $c_2$, but their two-page paper already contains a lot of physical insights as well. These include the interpretation of the vacuum as a nonlinear medium, the possibility of vacuum pair production and the smallness of the light-by-light scattering cross section in the optical regime which they stated as $\sigma \simeq 10^{-70}$ cm$^2$.

\section{Historical Interlude}\label{sect:HIST}

\begin{figure}[h]
\begin{center}
  \includegraphics[width = 0.35\textwidth]{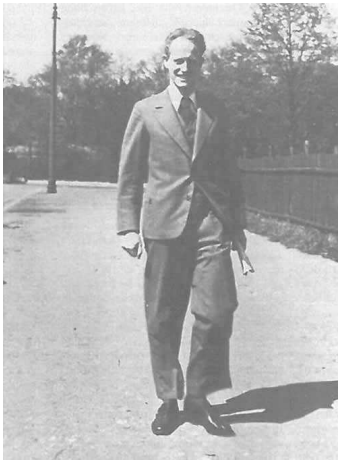}
\end{center}
\caption{\label{fig:Euler} Hans Euler ca.\ 1935. Picture: H.\ Wergeland (from Ref.~\protect\refcite{Hoffmann:1989}).}
\end{figure}

As this talk is a contribution to the session celebrating the 75th anniversary of the seminal paper ``Consequences of Dirac's theory of the positron'' by W.~Heisenberg and H.~Euler\cite{Heisenberg:1935qt} ist seems appropriate to devote a section on some of the history. As most of this has already been covered in Gerald Dunne's presentation\cite{Dunne:2011} I have decided to put my focus on one particular aspect: this year we commemorate not only the Heisenberg-Euler paper but also the 70th anniversary of Hans Euler's death. So let me say a few words about Heisenberg's student and co-author excerpting from the German article Ref.~\refcite{Hoffmann:1989} where also Euler's photograph\footnote{After my talk Iver Brevik, who was in attendance, kindly informed me that the photographer Harald Wergeland, a student friend of Euler's in Leipzig, was later to become one of Brevik's teachers in Norway.} (Fig.~\ref{fig:Euler}) is taken from.

Hans Heinrich Euler was born on 6 October 1909 in Meran, South Tyrol (Alto Adige), then part of the Austro-Hungarian Empire, now the northernmost province of Italy. In 1914 his family  moved to Germany (his father's home country) where Euler spent his undergraduate years in Munich, Bonn, G\"ottingen and Frankfurt. For the academic year 1933/34 he transferred to Leipzig where he became a PhD student of Heisenberg's (his supervisor being only 8 years his senior!). In the autumn of 1935 he submitted his thesis ``On the scattering of light by light according to Dirac's theory'' which was published in 1936\cite{Euler:1936}. On 25 June 1936 he was awarded his PhD with a mark of ``very good'' (from both Heisenberg and his second supervisor, F.~Hund). In the autumn of 1937 Heisenberg was able to provide Euler with a position as a ``subsidiary'' research assistant -- after a year or so of financial insecurity. As early as June 1938 Euler was able to submit his ``habilitation thesis'' on cosmic rays whereupon his position was upgraded to a regular assistantship. During the first year of World War II Euler was exempted from serving in the military. In 1940, though, apparently after a phase of depression, Euler volunteered for the Air Force -- despite his left leaning convictions. He was then trained as an on-board navigator and meteorologist. On 23rd June 1941 he was lost in action after his reconnaissance plane was shot down over the Sea of Azov, Crimea\cite{Hoffmann:1989}. The mystery of his ``disappearance'' is aggravated by contradicting historical sources. The German Wikipedia entry for Hans Euler, e.g.\ talks of an emergency landing due to engine failure on 24th June 1941 and his capture by fishermen the following day. This would suggest that Euler perished as a Soviet prisoner of war. There is even some rather apocryphal evidence that a notebook of Euler's containing calculations on Uranium chain reactions was found in early spring 1942 near Taganrog on the Azov Sea shore\cite{Sutyagin:1996}. It thus seems as if there is some work left for historians interested in the early history of QED\cite{Miller:1995} and the physicists involved.

In his thesis\cite{Euler:1936} Euler gave a detailed account of the Euler-Kockel synopsis of Ref.~\refcite{Euler:1935zz}, i.e.\ the determination of the leading order correction to the Maxwell Lagrangian.  In his own words, this correction represents ``an interaction between light quanta indicating the virtual production of matter...''. The extension of this result to \emph{all} orders in the external field was then given in the seminal Heisenberg-Euler paper\cite{Heisenberg:1935qt} we are celebrating in this conference session. As Gerald Dunne has nicely reviewed this paper anticipated many modern developments, among them subtraction techniques of UV renormalisation, the (implicit) evaluation of background determinants and the concept of low-energy effective field theory. In the next section I want to focus on another highlight of their results, the prediction and interpretation of vacuum polarisation of which they said: ``...even in situations where the [photon] energy is not sufficient for matter production, its virtual possibility will result in a `polarisation of the vacuum' and hence in an alteration of Maxwell's equations''\cite{Heisenberg:1935qt}.

\section{Strong Laser Fields: Examples}

\begin{figure}[h]
\begin{center}
  \includegraphics[width = 0.9\textwidth]{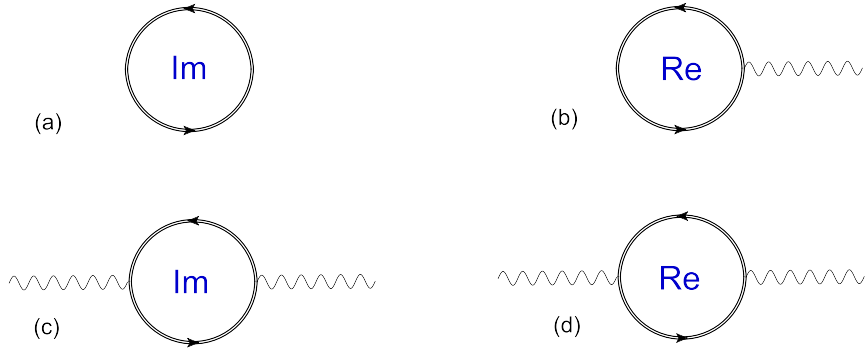}
\end{center}
\caption{Overview of different vacuum polarisation graphs.\label{fig:VacPolOverview}}
\end{figure}

We have seen above (Fig.~\ref{fig:VacPolExpanded}) that the standard QED vacuum polarisation of Fig.~\ref{fig:VacPol} becomes more complicated and hence describes a larger amount of physics when an external field is present. Even at lowest nontrivial order (altogether four external photon lines) there are several different cases depending on the number of non-laser photons involved.  By the optical theorem, each of these graphs may have an imaginary part corresponding to a particular variant of pair creation. In what follows we will discuss four examples: vacuum pair production, vacuum emission, stimulated pair production and vacuum birefringence. The associated Feynman diagrams (to all orders in laser photons) are displayed in Fig.~\ref{fig:VacPolOverview} (a)--(d), respectively.

\subsection{Vacuum pair production}

\begin{figure}[h]
\begin{center}
  \includegraphics[scale=1.2]{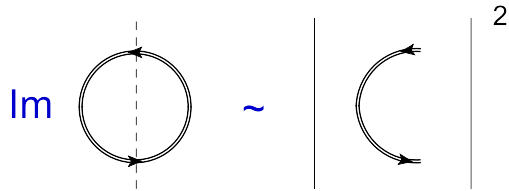}
\end{center}
\caption{Vacuum pair production via the optical theorem.\label{fig:OptThmVacPP}}
\end{figure}

For constant electric fields this is the celebrated Sauter-Schwinger mechanism\cite{Sauter:1931zz,Schwinger:1951nm} represented by Fig.~\ref{fig:VacPolOverview}~(a) which, via the optical theorem, yields pair production as shown in Fig.~\ref{fig:OptThmVacPP}.

The rate becomes substantial whenever the following invariant inequality holds,
\be
  E_0 \equiv  \left( \sqrt{\mathcal{S}^2 + \mathcal{P}^2} + \mathcal{S} \right)^{1/2} \gtrsim E_S \; .
\ee
For small (i.e.\ currently available) field strengths, $E_0 \ll E_S$, the rate is exponentially suppressed, $\mathfrak{R} \sim \exp(-\pi E_S/E_0)$. For null fields such as plane waves, however, the rate vanishes identically with the invariants $\mathcal{S}$, $\mathcal{P}$ and hence $E_0$. Thus, with laser fields, one has to fight \emph{both} the null field character \emph{and} the exponential suppression which makes vacuum pair production in a laser background a very difficult task. One can estimate the rate for Gaussian beams with geometry parameter $\kappa$ from (\ref{KAPPA}) as
\be
  \mathfrak{R} \sim  \kappa^2 \exp(-\pi E_S/ \kappa E_0) \; .
\ee
Hence, compared to a constant electric field one has additional suppression by powers of $\kappa \le 1/2\pi$ both in the prefactor and in the exponent. Using superpositions of beams such as standing waves the situation may be a bit more favourable\cite{Bulanov:2010ei}.

\subsection{Light-by-light scattering (vacuum emission)}

\begin{figure}[h]
\begin{center}
  \includegraphics[scale=0.9]{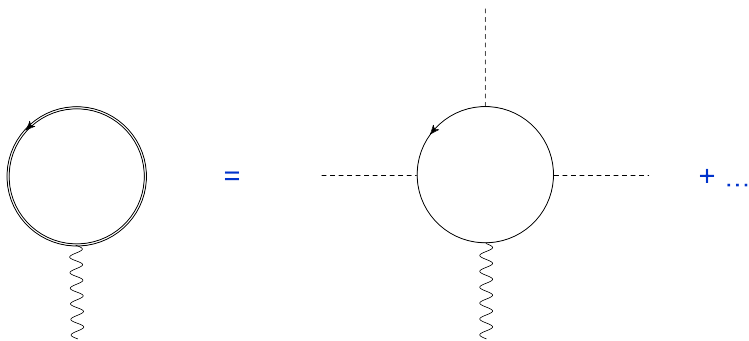}
\end{center}
\caption{Vacuum emission aka light-by-light scattering aka 3-wave mixing as suggested in \protect\refcite{Lundstrom:2005za}.\label{fig:VacEmission}}
\end{figure}

The scattering of light by light was originally predicted by Halpern\cite{Halpern:1934} in 1934 and shortly afterwards worked out in detail bei Euler, Heisenberg and Kockel as was discussed at length in Section~\ref{sect:HIST}. However, with all four external photons being real, this process has has never been observed in the lab. So, surprisingly, there remains a fundamental QED prediction of purely quantum nature that is still waiting for experimental confirmation\footnote{One should, however, note that the variant of Delbr\"uck scattering (two photons virtual) has been observed\cite{Jarlskog:1974tx,Schumacher:1975kv}. In addition, light-by-light scattering contributes to the electron anomalous magnetic moment $g-2$ at three-loop order, and in this sense has been tested, albeit in an indirect manner. For real $\gamma$-$\gamma$ scattering there are only bounds\cite{Bernard:2010dx}.}. The reason for this, in a sense, has two facets. In the MeV regime, the cross section is of typical QED size, about $10^{-30}$ cm$^2$, but in this regime one is lacking photon flux. On the other hand, as we have seen, in the optical regime the cross section is tiny, and so far fluxes have not been large enough to counterbalance this. Another simple estimate, however, shows that this situation may change in the near future. The crucial point is to utilise the huge photon densities provided by high-power lasers and employ the tadpole type graph Fig.~\ref{fig:VacPolOverview}~(b) which is expanded to lowest order in Fig.~\ref{fig:VacEmission}.

One should point out, though, that for null fields, the tadpole is zero as one cannot avoid contracting two null background field tensors, $\mathcal{S} \sim FF = 0$, in a fourth order expression with only a single non-laser field, $f$ [e.g.\ $(fF)(FF) = 0$]. For non-null configurations, the diagram survives, and the number of emitted photons may be estimated as follows.

The Euler-Kockel result\cite{Euler:1935zz} for the cross section is nowadays text book lore\cite{Jauch:1976} and reads
\be
  \sigma_{\gamma\gamma}  = \frac{973}{10125 \pi^2} \,  \alpha^2 \, r_e^2 \, \nu_L^6 \simeq 10^{-67} \, \mbox{cm}^2 \; ,
\ee
with fine structure constant $\alpha$, the classical electron radius, $r_e = \alpha/m$, dimensionless laser frequency $\nu_L$ as defined in (\ref{NU}) and the numerical value resulting for $\nu_L \simeq 10^{-6}$ (optical regime). The definition (\ref{A0}) implies a laser photon density
\be
  n_L \simeq 10^{14} \, a_0^2/ \mu\mbox{m}^3 \; ,
\ee
and hence a photon number $N_\gamma = 10^{17} a_0^2$ in a typical focus volume of $(10 \, \mu\mbox{m})^3$. The number of emitted photons for a `moderate' $a_0$ value of $10^2$ will then be
\be \label{N.GAMMA}
  N_{\gamma'} \simeq \frac{\sigma_{\gamma\gamma}}{(10 \, \mu\mbox{m})^2} \, N_\gamma^3 \simeq 10^{-4} \; .
\ee
This seems tiny but, as the number of emission events scales like $a_0^6$, we only need an oder of magnitude increase in $a_0$ to about $10^3$ to compensate the smallness of the cross section. If we assume a configuration involving Gaussian beams there may be additional suppression factors given by powers of $\kappa$ so that, clearly, a more detailed analysis will be required. Nevertheless, an observation of light-by-light scattering may become a realistic option in the near future.

\subsection{Vacuum birefringence and stimulated pair production}

Let me finally discuss the vacuum polarisation diagrams of Fig.s \ref{fig:VacPolOverview}~(c) and (d). The real part describes dispersive effects on the propagation of probe photons (birefringence) and the imaginary part absorptive ones where photons `disappear' to `make room' for electron positron pairs. Even for null fields we now do get a contribution as we can contract $(fF)^2 \ne 0$. The effective Lagrangian may then be written as
\be
   \mathcal{L}_{\mathrm{eff}} = \mathcal{L}_0 + \Delta \mathcal{L}_{\mathrm{eff}} = \frac{1}{2} a_\mu \left( \Box g^{\mu\nu} + \Pi^{\mu\nu}[A] \right) a_\nu \; ,
\ee
where the polarisation tensor can be seen to have the factorised form
\be
  \Pi^{\mu\nu} (A;k) = C^{\mu\nu}_{\;\;\;\alpha\beta}(A) \, k^\alpha k^\beta \; .
\ee
Upon diagonalisation one finds \emph{two} nontrivial eigenvalues of the polarisation tensor $\Pi^{\mu\nu}$ which, in the constant field approximation (plane waves becoming crossed fields) take on the form $\Pi_\pm = c_\pm k_\mu T^{\mu\nu}k_\nu = c_\pm I_2$. We recognise the invariant (\ref{I2}) resurfacing and refrain from specifying the numerical coefficients $c_\pm$ which are given in terms of the Euler-Kockel coefficients of the effective Lagrangian. The two eigenvalues imply two dispersion relations that can be expressed wit the help of an effective metric\cite{Shore:2007um,Heinzl:2006pn},
\be
  k^2 - \Pi_\pm = (g^{\mu\nu} - c_\pm T^{\mu\nu}) k_\mu k_\nu =  0 \; .
\ee
This in turn gives rise to \emph{two} indices of refraction first obtained in the seminal thesis of Toll\cite{Toll:1952rq},
\be \label{N}
  n_\pm = 1 + \frac{\alpha}{45 \pi} (11 \pm 3) \epsilon^2 \; .
\ee
This is the phenomenon of (vacuum) birefringence induced by a strong external field. The obvious question then is whether the effect is measurable. A possible experiment to detect vacuum birefringence is as follows\cite{Heinzl:2006xc}. One sends a linearly polarised probe beam through a high intensity laser focus and measures the ensuing ellipticity of the outgoing beam which is proportional to the difference in indices squared, $\delta^2 \sim (n_+ - n_-)^2$. The effect (i.e.\ $\delta^2$) is maximised by choosing (i) an angle of 45 degrees between probe polarisation and electric field, (ii) a high probe frequency (X-ray regime, $\nu \simeq 10^{-2}$), (iii) a large Rayleigh length and (iv) laser intensity $\epsilon$ as large as possible. For a current Petawatt laser one has $\epsilon \simeq 10^{-4}$ and $\delta^2 \simeq 10^{-11}$ while for ELI one may expect $\epsilon \simeq 10^{-2}$ and $\delta^2 \simeq 10^{-5}$. Obviously, this requires extreme polarisation purity for the X-ray probe beam. Earlier this year, a record value of $2.5 \times 10^{-10}$ for polarisation purity has been reported\cite{Uschmann:2011}. Again we can state that an observation of vacuum birefringence seems to become feasible.

\begin{figure}[h]
\begin{center}
  \includegraphics[angle=270,width = 0.49\textwidth]{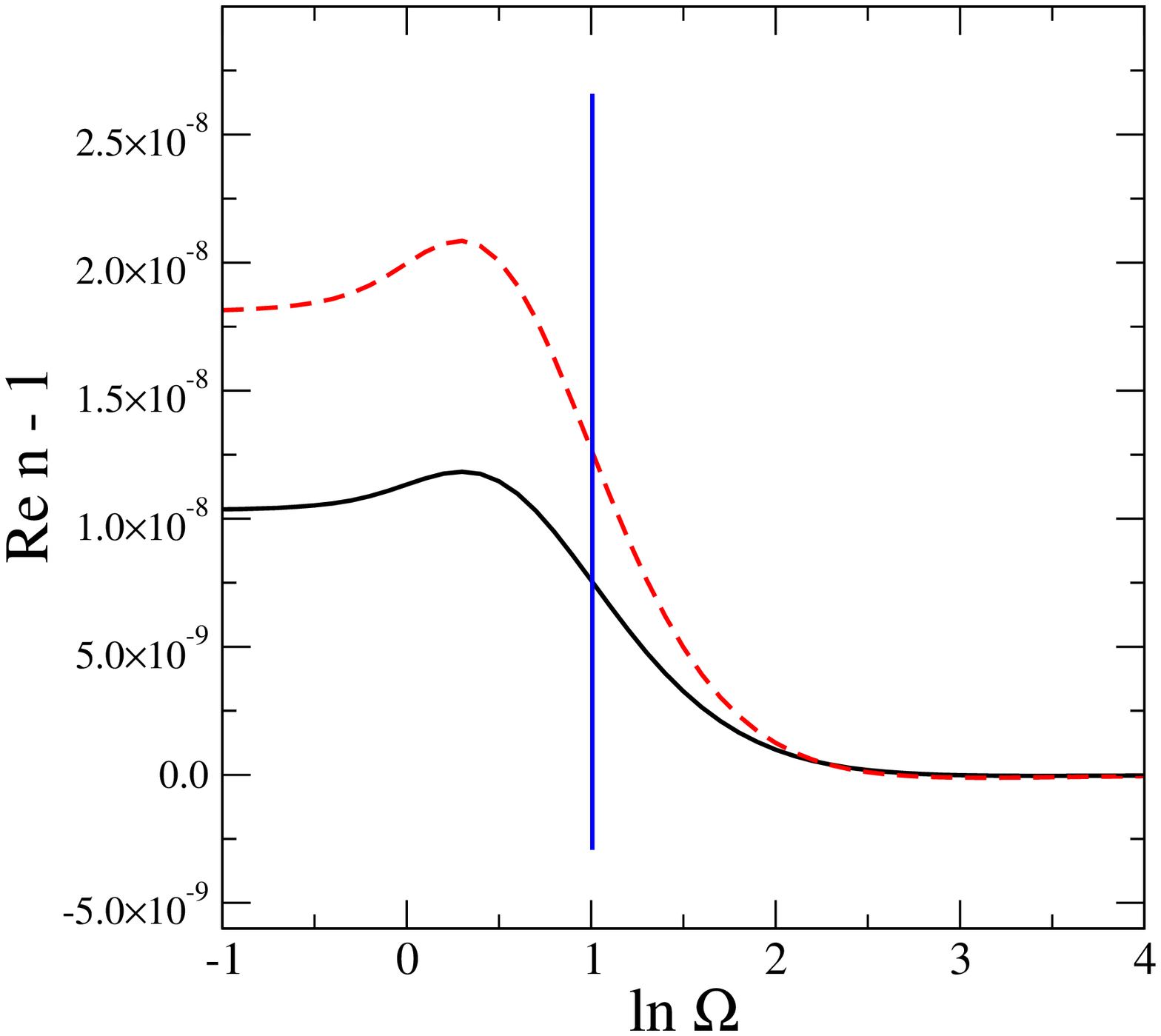}
  \includegraphics[angle=270,width = 0.49\textwidth]{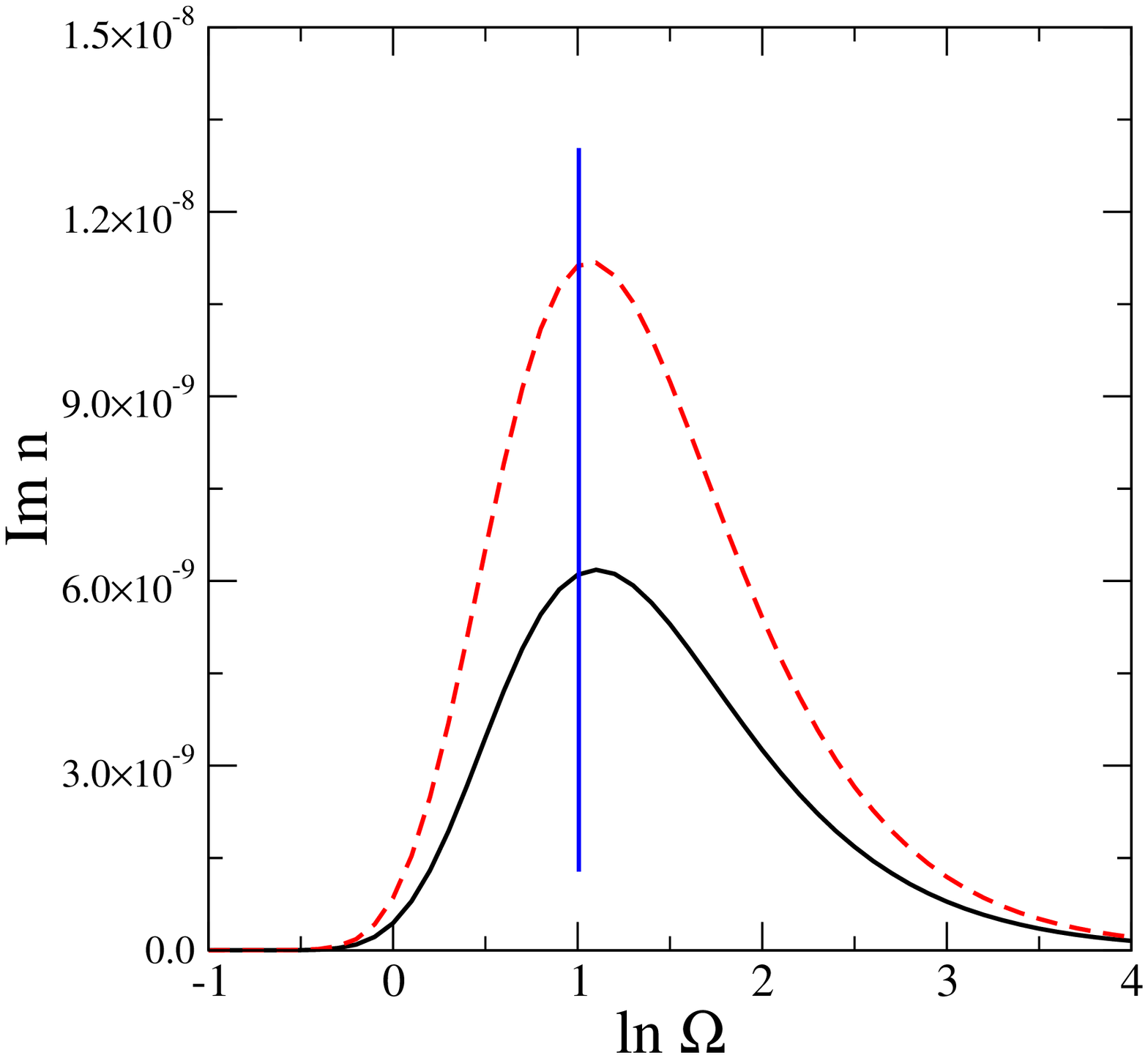}
\end{center}
\caption{Real and imaginary parts of the QED refractive indices $n_\pm$ as a function of $\ln \Omega \equiv \ln \epsilon\nu$. Dashed lines: $n_+$, full lines: $n_-$.\label{fig:N}}
\end{figure}

We conclude this section with an outlook towards higher probe energies $(\nu > 1)$. In this case the Heisenberg-Euler low energy approximation (valid for $\nu \ll 1$) breaks down and one has to determine the full polarisation tensor (to all orders in external momentum). For crossed fields this has already been achieved by Toll\cite{Toll:1952rq} (see  Ref.~\refcite{Dittrich:2000zu} for an overview). For plane wave backgrounds one has the more complicated results of Ref.s~\refcite{Becker:1975,Baier:1975ff}. We only consider the crossed field result which yields for both indices an expression of the form
\be
  n_\pm = 1 + \frac{2\alpha}{\pi}  \epsilon^2 \, f_\pm (\epsilon \nu) \; ,
\ee
with functions $f_\pm$ of the product $\epsilon\nu$ having both real and imaginary parts \footnote{In the limit $\nu \to 0$ one recovers (\ref{N}).}. These are displayed in Fig.~\ref{fig:N} assuming a scenario where $\epsilon\nu \simeq 3$ (blue vertical lines) is achieved by Compton backscattering off a few GeV electron beam. The important thing to note is that in this case one is well into the regime of anomalous dispersion ($dn/d\nu < 0$, left panel of Fig.~\ref{fig:N}) \emph{and} (via causality or a Kramers-Kronig relation\cite{Heinzl:2006pn}) the regime of absorption (Im $n \ne 0$, right panel of Fig.~\ref{fig:N}). Thus, if for some reason positrons cannot be observed, one may alternatively look for anomalous dispersion as a signal of pair production.

\section{Conclusion}

We have seen that laser power is approaching the Exawatt regime with unprecedented magnitudes in field strength and intensity. This corresponds to an uncharted region of the standard model, characterised by a peculiar external field that is not only ultra-strong but also \emph{pulsed, oscillatory and near null}.

The associated theory, a particular realisation of strong-field QED, has been developed long ago\cite{Toll:1952rq,Reiss:1962,Nikishov:1963,Brown:1964zz} but has never been tested for laser amplitudes $a_0 > 1$. In this high-intensity regime new theoretical challenges are to be met such as finite pulse duration and size effects which require going beyond the plane wave approximation. As intensities grow larger the issue of radiation backreaction will become more pressing (see e.g.\ Ref.s~\refcite{Harvey:2010ns,Harvey:2011dp} and references therein), and one may have to modify or even abandon the external field approximation.

With new high power laser facilities planned or under way\cite{Vulcan10PW:2009,ELI:2009}, the intensity frontier will be further pushed upwards in the near future to values of $a_0 \simeq 10^3$. In this regime light-by-light scattering of real (optical) photons may become observable for the first time. In addition, the observation of vacuum birefringence should be feasible. This is particularly interesting as this process represents a possible window to physics beyond the standard model (axions or other weakly interacting particles) as has been reviewed by Guido Zavattini at this conference\cite{Zavattini:2011} (see also Ref.~\refcite{Ahlers:2007qf} and references therein).

In any case it seems we are experiencing the advent of a new sub-field of physics, namely high-intensity QED.  Both the realisation of old predictions and the discovery of new intensity effects may be just around the corner.

\section*{Acknowledgments}

It is a pleasure to thank the organisers, in particular Manuel Asorey and Emilio Elizalde, for providing a stimulating conference atmosphere in quite breathtaking surroundings. I am indebted to C.~Harvey, A.~Ilderton and M.~Marklund for a fruitful collaboration on some of the topics covered.
%
%


\end{document}